\documentclass[letterpaper,twocolumn,prb,nofootinbib,superscriptaddress,showpacs]{revtex4}

\usepackage{amsmath}
\usepackage{amssymb}
\usepackage{graphicx}
\usepackage[usenames,dvipsnames]{xcolor}
\usepackage{booktabs}
\usepackage{todonotes}
\usepackage{microtype}
\usepackage{hyperref}
\definecolor{myblue}{HTML}{0074D9}
\definecolor{myorange}{HTML}{FF851B}
\definecolor{myteal}{HTML}{39CCCC}
\definecolor{myred}{HTML}{FF4136}
\definecolor{mygray}{HTML}{AAAAAA}
\definecolor{mysilver}{HTML}{EFEFEF}
\hypersetup{colorlinks=true,citecolor=myblue,linkcolor=myblue,urlcolor=myred}
\presetkeys{todonotes}{backgroundcolor=mysilver, bordercolor=white}{}
\tikzset{/tikz/notestyleraw/.append style={text=myred}}

\hypersetup{colorlinks=true,citecolor=Red,linkcolor=Blue,urlcolor=Black}
\AtBeginDocument{
	\heavyrulewidth=.08em
	\lightrulewidth=.05em
	\cmidrulewidth=.03em
	\belowrulesep=.65ex
	\belowbottomsep=0pt
	\aboverulesep=.4ex
	\abovetopsep=0pt
	\cmidrulesep=\doublerulesep
	\cmidrulekern=.5em
	\defaultaddspace=.5em
}

\begin{document}

	%%% ADDITION STARTS
	%\vspace*{\fill}
	%\begin{tabular}{p{0.06\textwidth}p{0.76\textwidth}p{0.18\textwidth}}
	%		&
	%		\fbox{\begin{tabular}{p{0.76\textwidth}}
	%						This manuscript has been authored by UT-Battelle, LLC
	%						under Contract No. DE-AC05-00OR22725 with the U.S. Department of Energy. The
	%						United States Government retains and the publisher, by accepting the article for
	%						publication, acknowledges that the United States Government
	%						retains a non-exclusive, paid-up, irrevocable, world-wide license
	%						to publish or reproduce the published form of this manuscript, or allow others to
	%						do so, for United States Government purposes. The Department of Energy will
	%						provide public access to these results of federally sponsored
	%						research in accordance with the DOE Public Access
	%						Plan(\nolinkurl{https://energy.gov/downloads/doe-public-access-plan} ).
	%				\end{tabular}}
	%	\end{tabular}
	%\vspace*{\fill}
%	\cleardoublepage
	%%% ADDITION ENDS
%
% Title Page
%
\title{Root-$N$ Krylov-space correction-vectors for spectral functions with the density matrix renormalization group}

\author{A.~Nocera}
\affiliation{Department of Physics and Astronomy and Stewart Blusson Quantum Matter Institute, University of British Columbia, Vancouver, B.C., Canada, V6T, 1Z4}

\author{G.~Alvarez}
\affiliation{Computational Sciences and Engineering Division,
 \mbox{Oak Ridge, Tennessee 37831}, USA}

\date{\today}
% FIX THE PACS
%\pacs{71.10.Fd, 71.27.+a, 02.70.Hm, 79.60.-i}
%\setlength{\intextsep}{10mm}
\begin{abstract}
We propose a method to compute spectral functions of generic Hamiltonians
using the density matrix renormalization group (DMRG) algorithm directly in the frequency domain,
based on a modified Krylov space decomposition to compute the correction-vectors.
Our approach entails the calculation of the root-$N$ ($N=2$ is the standard square root) of the Hamiltonian propagator
using Krylov space decomposition, and repeating this procedure $N$ times to obtain the actual correction-vector.
We show that our method greatly alleviates the burden of keeping a large bond dimension at large target frequencies, 
a problem found with conventional correction-vector DMRG,
while achieving better computational performance at large $N$.
We apply our method to spin and charge spectral functions of $t$-$J$ and Hubbard models
in the challenging two-leg ladder geometry, and provide evidence that the root-$N$ approach reaches a 
much improved resolution compared to conventional correction-vector.
\end{abstract}
\maketitle

\section{Introduction}
In condensed matter physics, several unusual properties of strongly correlated quantum materials are unveiled
using \emph{spectroscopic} techniques, such as angle resolved photoemission spectroscopy (ARPES)~\cite{re:Damascelli2003},
inelastic neutron scattering (INS), and resonant inelastic x-ray scattering (RIXS)~\cite{re:Ament2011}.
These experimental probes do not provide a direct access to the ground state, but
rather explore the low energy excitations of the system.
Excitations spectra are esperimentally measured looking at the energy and momentum
exchanged by the probe of each technique with the material:
photo-emitted electron for ARPES, neutron for INS, photon for RIXS,
and are theoretically encoded in \emph{spectral functions}.
The progressive improvement in momentum and energy resolution in experimental spectroscopic apparatus
calls on the theory side for an equally significant improvement of the spectral functions
calculations accuracy.

For a one-dimensional (1D) lattice Hamiltonian of size $L$, a generic spectral function can be defined as
\begin{equation}
	O(q,\omega) = \frac{1}{L}\sum_{i,j} e^{i q (i-j)} \int^{\infty}_{0} dt e^{i (\omega+E_g)t}\langle \psi| \hat{O}_{i} e^{-i\hat{H}t} \hat{O}_{j}|\psi\rangle,
\end{equation}
where $|\psi\rangle$ is the ground state of the system Hamiltonian $\hat{H}$,
$E_g$ is the ground state energy, $q$ and $\omega$ are the momentum
and frequency (or energy) of the electron in the material, and $\hat{O}_{j}$ is the relevant operator involved
in the scattering process of the specific technique acting locally on site ``$j$" ($\hat{O}_{j}=\hat{c}_{j\sigma}$ for ARPES,
$\hat{O}_{j}=\hat{S}^{z}_{j}$ for INS, while special care is needed for RIXS, as written in Ref.~\cite{re:Nocera2018}).

In 1D, the most powerful method to compute spectral functions of arbitrary strongly correlated Hamiltonians is the
density matrix renormalization group (DMRG)~\cite{re:White1992,re:White1993};
the DMRG is a variational but systematically exact algorithm to find a matrix product state (MPS) representation
for the ground state of the system~\cite{re:Schollwock2011}.
Spectral functions can be computed in the time-space domain using time dependent
matrix product state methods~\cite{re:Feiguin2004,re:Kollath2004}. (For a recent review of the different
variants, see Ref.~\cite{re:Paeckel2019})
When using time evolution, the problem is to find an efficient MPS representation of the time-evolved vector
\begin{equation}
	|x_j(t)\rangle=e^{-i \hat{H} t}\hat{O}_{j}|\psi\rangle,
\end{equation}
where the ground state of the Hamiltonian $\hat{H}$ is locally modified by the $\hat{O}_{j}$,
and the resulting state is evolved up to a very large (in principle ``infinite'') time.
This evolution always grows the entanglement of the state, and thus spoils the compression of the MPS representation.
Simulations are therefore typically stopped
at some large or maximum time, and linear prediction \cite{re:White2008} or recursion methods \cite{re:Tian2021}
are needed
to obtain a well behaved Fourier transform in frequency.

In this paper, we are concerned with the complementary approach of computing the spectral functions
\emph{directly in the frequency domain.}
To discuss this case, it helps to rewrite the spectral function as
\begin{align}
	O(q,\omega) &= \lim_{\eta\rightarrow0}\frac{1}{L}\sum_{i,j} e^{i q (i-j)}\times\nonumber\\
	&\times -\frac{1}{\pi}\textrm{Im}\Bigg[\langle \psi| \hat{O}_{i} \frac{1}{\omega-\hat{H}+E_g+i\eta}\hat{O}_{j}|\psi\rangle\Bigg],
\end{align}
where one writes down the Hamiltonian propagator explicitly, and $\eta>0$ is an arbitrary small extrinsic spectral broadening.
Three are the approaches that are typically used by DMRG practitioners.
Historically, the hybrid DMRG-Lanczos-vector methods were first introduced \cite{re:Hallberg1995}
(refined using MPS more recently \cite{re:Dargel2011,re:Dargel2012}),
then afterwards the correction-vector (CV) method\cite{re:Kuhner1999,re:Pati1997,re:Jeckelmann2002,re:Jeckelmann2008,re:Weichselbaum2009,re:NoceraPRE2016}
and Chebyshev polynomial methods \cite{re:Holzner2011,re:Wolf2015,re:Halimeh2015,re:Xie2018} were proposed.
In the CV method, one computes the real and imaginary part of the correction-vector
\begin{equation}
	|x_j(\omega+i\eta)\rangle=\frac{1}{\omega-\hat{H}+E_g+i\eta}\hat{O}_{j}|\psi\rangle
\end{equation}
at fixed frequency $\omega$, finite broadening $\eta$, and then computes the spectral function
in real-frequency space as a stardard overlap $\langle \psi| \hat{O}_{i}|x_j(\omega+i\eta)\rangle$.
The real and imaginary part of the correction-vector
are typically obtained by solving for coupled matrix equations using conjugate-gradient methods~\cite{re:Kuhner1999},
or by minimizing a properly defined functional~\cite{re:Jeckelmann2002,re:Jeckelmann2008}. 
Ref.~\cite{re:Weichselbaum2009} formulated the algorithm in MPS language.

In 2016, we proposed~\cite{re:NoceraPRE2016} an alternative method to compute
directly the correction-vectors using a Krylov space expansion of the Hamiltonian operator
constructed starting from the locally modified MPS $|\phi\rangle=\hat{O}_{j}|\psi\rangle$.
In all these cases, the entanglement content of the correction-vectors is large, and it can
be very large for large frequencies.
This makes standard CV DMRG simulations very expensive
for Hamiltonians beyond spin systems or for large lattices.

In 2011, Holzner et al.~\cite{re:Holzner2011} proposed a MPS method to compute
a Chebyshev polynomial expansion (truncated at some order $N$) of the spectral function (CheMPS).
In this approach, the Chebyshev momenta can be obtained from overlaps of a properly defined series of Chebyshev vectors.
The main advantage of CheMPS lies in the small entanglement that each Chebyshev vector has,
because the method \emph{redistributes} the large entanglement
of the correction-vectors $|x_j(\omega+i\eta)\rangle$ for different frequencies
(or alternatively the time-evolved state $|x_j(t)\rangle$) over the entire series of Chebyshev vectors.

Inspired by this idea, we here propose a method to compute
a generalized correction-vector with smaller entanglement content, the root-$N$ correction vector, defined as
\begin{equation}\label{eq:rootNn1}
|x^{1/N}_j(\omega+i\eta)\rangle=\Bigg(\frac{1}{\omega-\hat{H}+E_g+i\eta}\Bigg)^{1/N}\hat{O}_{j}|\psi\rangle.
\end{equation}
The idea is to construct the actual correction-vector as the final vector
of the series $\{|x^{p/N}_j(\omega+i\eta)\rangle\}_{p\in[1,N]}$
after $N$ applications of the root-$N$ propagator.
At first sight, it seems that, if $N$ is sufficiently large, constructing the entire series of vectors
just adds a computational overhead compared to the standard DMRG CV algorithm,
because only the final vector of the series is actually needed for the spectral function calculation.
Yet we will show that the entanglement content of the series slowly builds up with $p$,
and therefore, going through many intermediate steps
is more efficient than the conventional DMRG CV algorithm, which tries
to compute the last element of the series in one step only.

The paper is organized as follows. Section II.A introduces the main steps of the algorithm;
section II.B analyzes the algorithm's computational performance, and the entanglement content
of the root-$N$ correction-vectors in the test case of a Heisenberg model in the two-leg ladder geometry.
Section II.C applies our root-$N$ method to compute spin and charge spectral functions of doped
$t$-$J$ and Hubbard models in the challenging two-leg ladder geometry,
showing how our method improves the energy resolution at large frequencies.
Finally, we present our conclusions and outlook.

\section{Method and Results}
\subsection{root-$N$ CV method algorithm}

The algorithm follows four steps. We assume a standard DMRG approach but provide the main
step of the algorithm in MPS language in Appendix~\ref{appendix_MPSalgo}.
\begin{enumerate}
	\item [1.] Compute the ground state wave function with the DMRG.
\end{enumerate}
		For each frequency $\omega,$ repeat the steps 2--4 to cover the desired interval $[\omega_{\text{min}},\omega_{\text{max}}]$
		with some step $\Delta\omega>0$:
\begin{enumerate}
	\item[2.] Apply the operator $O_j$ at the center of the chain and build the $p=1$ root-$N$ correction vector
		$|x^{p/N}_j(\omega+i\eta)\rangle$ as in Eq.~(\ref{eq:rootNn1}).
		This can be done using conventional DMRG as described in Ref.~\cite{re:NoceraPRE2016}.
		Appendix~\ref{appendix_MPSalgo} describes in detail the algorithm in MPS language.
		In this stage, as in the conventional CV method, the sources of error are two:
		the Lanczos error in the tridiagonal decomposition of the Hamiltonian (or effective Hamiltonian in MPS language);
		the SVD error of the multi-targeting DMRG procedure (state-averaging in MPS language).
\end{enumerate}
		Repeat step 3 until the $N$th root-$N$ correction vector is constructed and optimized, then go to step 4.
\begin{enumerate}
	\item[3.] Build the $p$$+$$1$ root-N correction vector $|x^{(p+1)/N}_j(\omega+i\eta)\rangle$  from the previous one assuming
		it as a starting point for the Krylov space decomposition of the Hamiltonian.
		A few DMRG sweeps are performed until a desired convergence is reached.

	\item[4.] Measure the spectral function in real-frequency space as the overlap
		$\langle \psi| \hat{O}_{i}|x_j(\omega+i\eta)\rangle$; this part is the same as in conventional DMRG CV.
\end{enumerate}

\begin{figure*}
\centering
\begin{minipage}{.425\textwidth}
  \centering
  \includegraphics[width=\columnwidth]{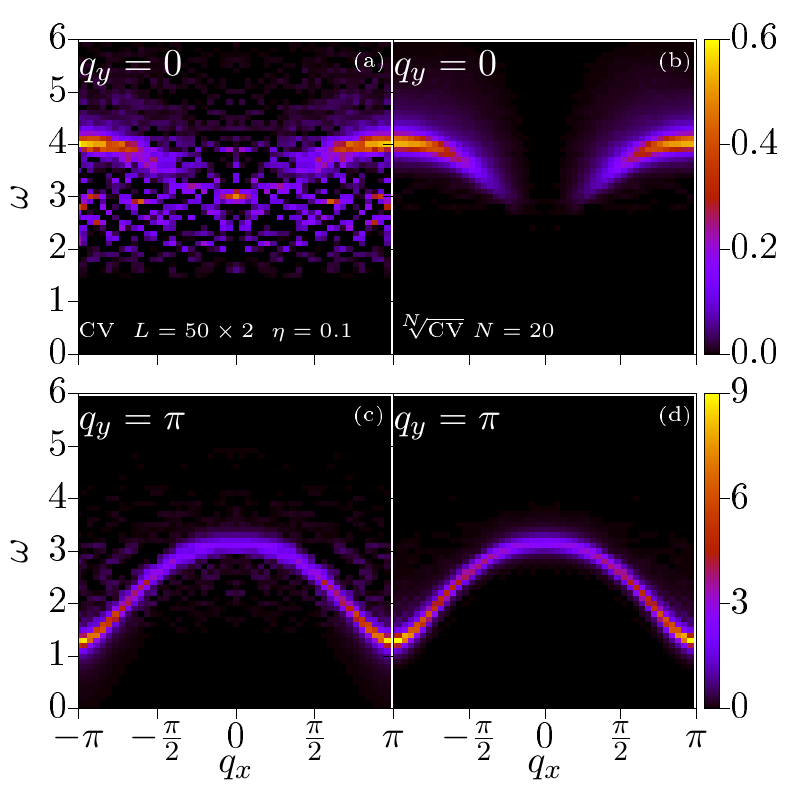}
\end{minipage}%
\begin{minipage}{.475\textwidth}
  \centering
  \includegraphics[width=\columnwidth]{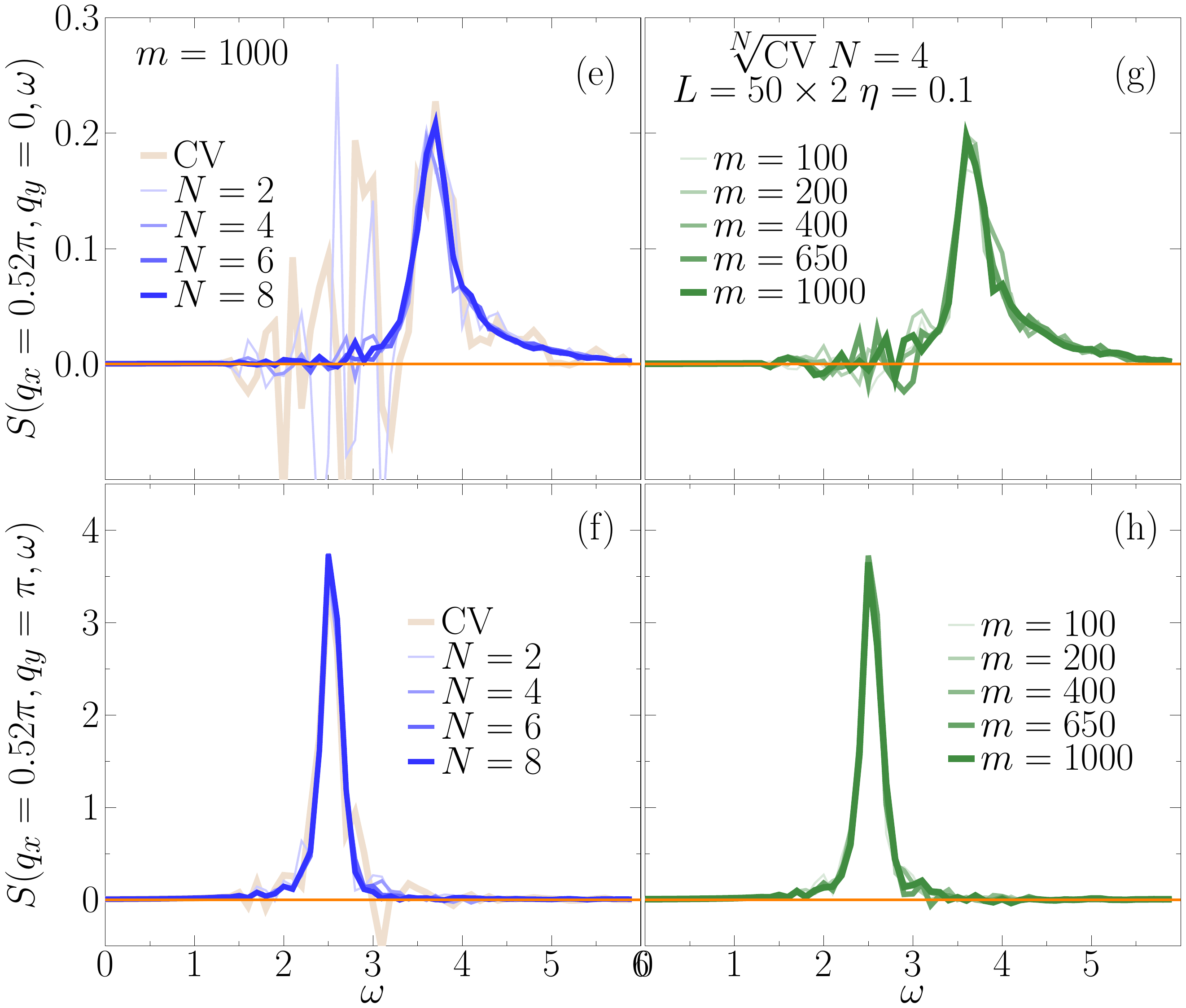}
\end{minipage}
	\caption{\textbf{Convergence analysis of the root-$N$ Krylov space Correction-Vector (CV) DMRG method: the Heisenberg ladder.}
        Panels (a)-(b) report the $q_y=0,\pi$ components of the $S(q_x,q_y,\omega)$
        using the standard Krylov space CV approach.
        A ladder with $J_y=2J_x$ is simulated. Length is $L=50\times2$, broadening $\eta=0.1$,
        and resolution step $\delta\omega=0.1$ (units are set by $J_x=1$).
        Panels (c)-(d) report the $q_y=0,\pi$ components of the $S(q_x,q_y,\omega)$ using the root-$N$
        Krylov space CV method using $N=20$.
	Panels (e)-(f) show specific momentum-energy line cuts of the dynamical spin structure computed in panels (a)-(d)
	for different values of the \emph{exponent} $N$.
        Numerical fluctuations and instabilities are removed, and quality of the spectra is clearly improved as $N$ is increased.
	Panels (g)-(h) show the same line cuts as in Panels (e)-(f) but with at fixed $N$, and increasing the number of DMRG states.
	When using the root-$N$ CV method with $N=4$, a substantially smaller number of states $m=200-650<1000$ suffices to get
        better quality results than with the standard Krylov space DMRG CV approach.}
        \label{fig:1}
\end{figure*}

To clarify the main steps of the algorithm, we draw an analogy with Krylov space
approaches for time evolution of MPSs.
In this case, one constructs the time-evolved vector $|x_j(\delta t)\rangle=U(\delta t)\hat{O}_{j}|\psi\rangle$
only for \emph{small} time step intervals of length $\delta t=t/N$, and where $U(\delta t)=e^{-i \hat{H} \delta t}$
is the time evolution propagator. To get the final time-evolved vector at time $t$, one repeatedly applies $U(\delta t)$ to the MPS.
In practice one does not build the evolution operator $U(\delta t)$ in the local basis but
rather directly construct the vector $|x_j(\delta t)\rangle=U(\delta t)\hat{O}_{j}|\psi\rangle$ using
a Krylov space decomposition of the Hamiltonian (or effective Hamiltonian in MPS language).

In our proposed root-$N$ Krylov space approach, we introduce a propagator in a fictitious time space $s$
as $|x_j(\delta s)\rangle=W(\delta s)\hat{O}_{j}|\psi\rangle$, where $\delta s=1/N$, and where $W(\delta s)=e^{-\hat{K}\delta s}$,
with $\hat{K}=\ln{[\omega+E_g+i\eta-\hat{H}]}$. Clearly, if we apply $W(\delta s)$
$N$ times to the initial vector $\hat{O}_{j}|\psi\rangle$ we obtain the desired standard correction vector.
In other words, we \emph{formally} define and solve an auxiliary differential equation
\begin{equation}
        \frac{d}{d s}|x^s_j(\omega+i\eta)\rangle  = -\ln{[\omega+E_g+i\eta-\hat{H}]} |x^s_j(\omega+i\eta)\rangle,
\end{equation}
such that at time $s$ we have the solution
\begin{equation}
        |x^s_j(\omega+i\eta)\rangle  = \Bigg(\frac{1}{\omega-\hat{H}+E_g+i\eta}\Bigg)^{s} |x^0_j(\omega+i\eta)\rangle,
\end{equation}
with the initial condition being $|x^0_j(\omega+i\eta)\rangle = \hat{O}_{j}|\psi\rangle$.

In this construction, $1/N$ plays the role of a \emph{small} parameter to compute the resolvent
in the standard CV approach.
We will show that this method is especially useful at large target frequencies, where large bond
dimensions (or DMRG states) are typically needed.

\begin{figure*}
        \centering
        \includegraphics[width=2\columnwidth]{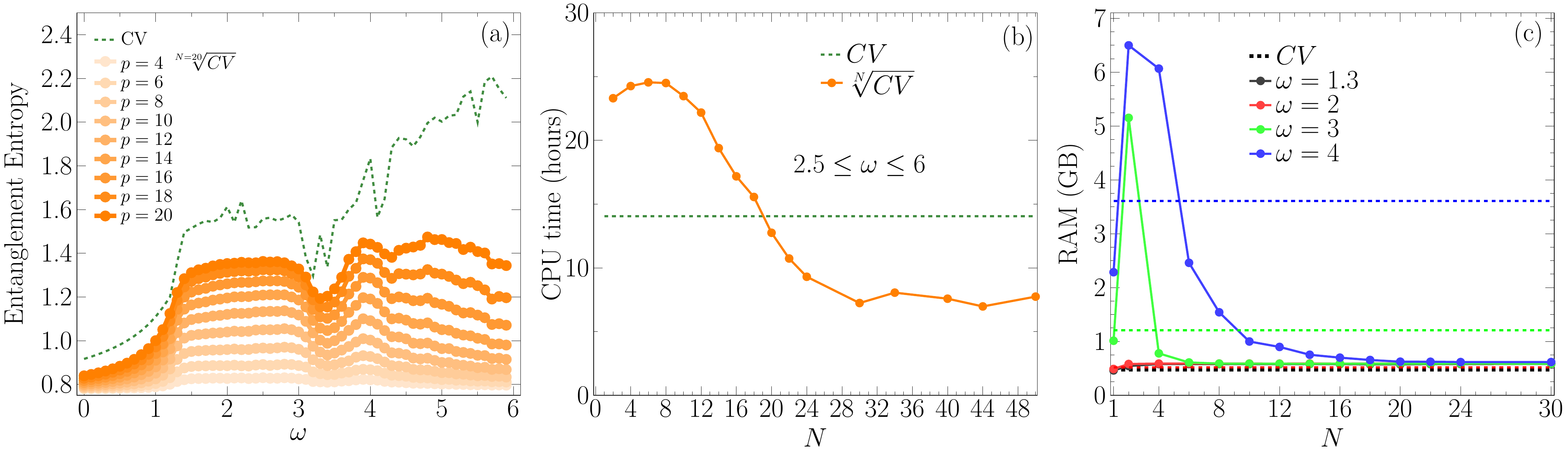}
        \caption{\textbf{Entanglement Entropy of the root-$N$ Krylov correction vectors and computational performance of the method.}
        Panel (a): Entanglement Entropy computed with conventional CV (dark green) compared
        to the same quantity computed with the root-$N$ CV; $N=20$ and different $p$ values (indicated by orange lines, 
	and thickness increasing with $p$),
        as a function of $\omega$. We have used a Heisenberg two-leg ladder with $J_y=2J_x$,
        length $L=50\times2$, as in Fig.~\ref{fig:1}.
        (b) Total CPU time in hours (obtained summing all the CPU times of the CV
        simulations in the frequency interval investigated) and (c) RAM
        as a function of the \emph{exponent} $N$. The simulations were run on a single Intel Xeon Gold 6130 CPU node
        (using hyperthreading over 8 cores).
        When compared with the standard Krylov space Correction-Vector method, the performance of the root-$N$ approach
        is superior for either sufficiently small or sufficiently large $N$.}
        \label{fig:2}
\end{figure*}

\subsection{Convergence analysis and Computational performance: the Heisenberg model on a two-leg ladder as a case study}

We begin by testing our root-$N$ CV method by applying it
to an isotropic Heisenberg model on a two-leg ladder geometry.
(The supplemental material~\cite{supplemental} provides computational details for all the models considered in this work.)
The Heisenberg Hamiltonian is the standard one,
with antiferromagnetic exchange interactions along both the leg and rung directions, and with
$J_y=2J_x$; open boundary conditions along the leg direction are assumed.
The explicit Hamiltonian expression is given in Eq.~\ref{eq:Heis} of Appendix~\ref{appendix_Hamis}.

Figure~\ref{fig:1}a-b reports spectral maps of the two components $q_y=0,\pi$
of the dynamical spin structure factor $S(\textbf{q},\omega)$
as a function of the momentum transfer $q_x$ along the leg direction, and of the frequency. (Definition of the spin structure factor is provided in Eq.~\ref{eq:Sqw} of Appendix~\ref{appendix_Hamis}.)
These are obtained with conventional CV as in Ref.~\cite{re:NoceraPRE2016} on a system size
of length $L=50\times 2$ and with an extrinsic broadening parameter $\eta=0.1J_x$.
By comparison, Fig.~\ref{fig:1}c-d reports results obtained using the root-$N$ CV method with $N=8$.
In both cases, we have used a maximum $m=m_{\text{max}}=1000$ DMRG states and a minimum $m_{\text{min}}=200$,
keeping the truncation error below $10^{-7}$.
(Please see the description around Eq.~(\ref{eq:App}) in Appendix~\ref{appendix_MPSalgo} for the definition of
the extended MPS which is optimized by SVD in the root-$N$ CV algorithm.)
We clearly notice an overall improved
spectrum in this case with respect to the conventional CV method.
We analyze below the spectral features in more detail.

Figure~\ref{fig:1}e-f shows momentum $q_x=0.52\pi$ line cuts of the spin spectra for the
$q_y=0,\pi$ components in the root-$N$ CV method. The data
shows that by increasing the \emph{exponent} $N$ numerical fluctuations,
instabilities are removed with respect to the conventional CV results.
The red curve in Fig.~\ref{fig:1}e-f shows that the conventional CV approach can yield
negative values for certain frequencies.
As finite size effects are small for a $L=50\times2$ ladder, these are clearly artifacts of the CV method
which might spoil important properties of the spectral functions such as sum rules.
On the contrary, the root-$N$ CV approach shows always positive values which
progressively improve upon increasing the exponent $N$.
Figure~\ref{fig:1}g-h shows how well the root-$N$ method converges with respect to the number of DMRG states.
Contrary to panels (a)-(d), in these panels the data for $m<1000$ was obtained
by imposing zero truncation error in the DMRG SVDs, therefore
setting $m=m_{\text{max}}=m_{\text{min}}$.
Our data shows that at fixed exponent $N=4$, a substantially smaller number of DMRG states $m=200-650<1000$
is sufficient to get better quality results than with the conventional CV approach. As we will show next,
this improvement can be understood by the much lower entanglement content of the root-$N$ correction-vectors.

Figure~\ref{fig:2}a shows indeed that the entanglement content of the root-$N$ correction vectors is smaller than the actual (conventional)
correction-vector. In this calculation, to compute the entanglement entropy of the expanded MPS for root-$N$ correction vectors
(Appendix Eq.~(\ref{eq:App}) has the definition), we
have used a maximum $m=2000$ DMRG states (and a minimum $m_{\text{min}}=200$),
keeping the truncation error below $10^{-8}$ in both methods. It is nice to see that the entanglement entropy
of the extended MPS in the root-$N$ CV method is very close to that of the conventional CV
in the lower frequency range investigated $\omega\in[0,\omega^{*}]$, with $\omega^{*}\simeq4.5$.
For larger frequencies, the root-$N$ CV approach
truncates the entanglement contained in the conventional CV vector,
showing that a larger exponent $N$ or a larger number of DMRG states should be considered.
Yet we highlight that this truncation does not show instablibilities or fluctuations as in the conventional CV approach.

In the same range of frequencies, we have monitored the CPU time taken for the simulations to complete in the two methods.
For moderately large exponent $N$, the root-$N$ CV method and the conventional CV method
have similar performances for sufficiently small exponent $N$, as can be seen in the dashed and solid circle lines
of Fig.~\ref{fig:2}b.
Eventually, when the entanglement is decomposed in smaller chunks by considering a larger $N$, the root-$N$ method becomes faster
even though many more optimizations and Lanczos decompositions are actually performed.
Figure~\ref{fig:2}c ends this subsection by showing further how memory requirements decrease at a large exponent $N$, as larger 
and larger frequencies are targeted.

\begin{figure*}
\centering
\begin{minipage}{.45\textwidth}
  \centering
  \includegraphics[width=\columnwidth]{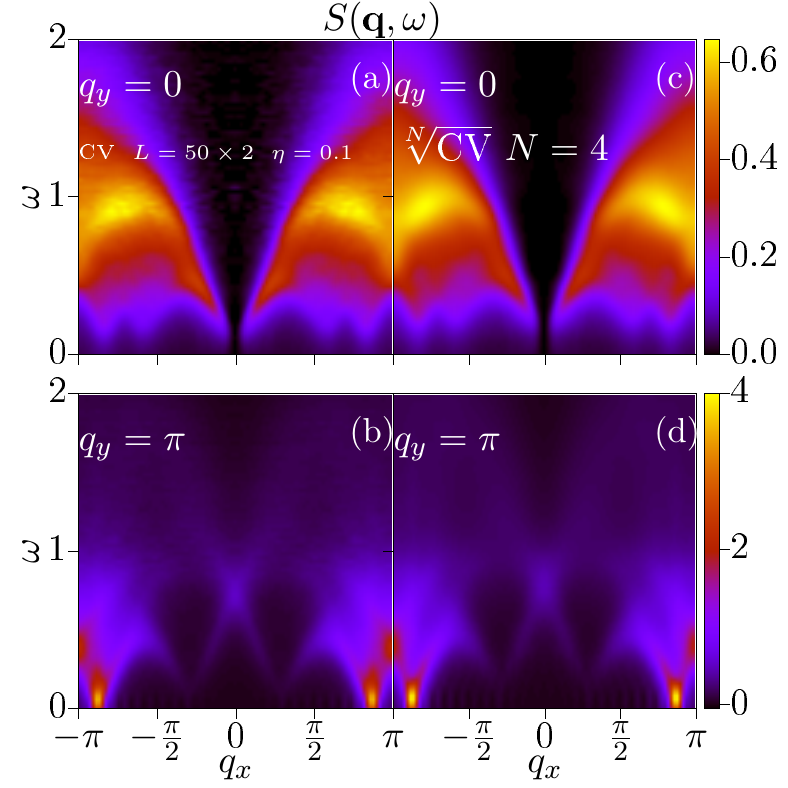}
\end{minipage}%
\begin{minipage}{.45\textwidth}
  \centering
  \includegraphics[width=\columnwidth]{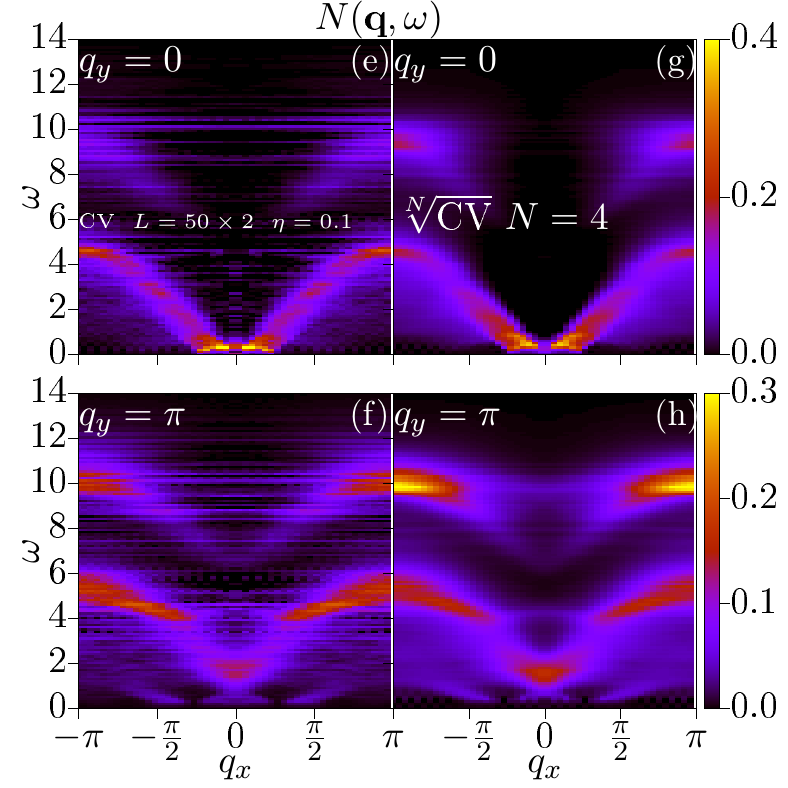}
\end{minipage}
        \caption{\textbf{Dynamical structure factors for a Hubbard two-leg ladder.}
        The $q_y=0$ and $q_y=\pi$ components of the dynamical spin structure factor $S(q_x,q_y,\omega)$ 
        using, in (a)-(b) the conventional Krylov space DMRG CV approach, against in (c)-(d) the root-$N$
	Krylov space CV method with $N=4$ (a maximum of $m=3000$ DMRG states were used.)
        The $q_y=0$ and $q_y=\pi$ components of the dynamical  charge correlation $N(q_x,q_y,\omega)$, 
        using, in (e)-(f) the conventional Krylov space DMRG CV, against in (g)-(h) the root-$N$
        Krylov space CV with $N=4$ (a maximum of $m=2000$ DMRG states were used.).
        A ladder $L=50\times2$ ladder with $t_y=t_x=t=1$, $U=8t$ is simulated with $N_{\text{el}}=0.88L$ electrons,
        broadening $\eta=0.1$,
        and resolution step $\delta\omega=0.025$ for $S(q_x,q_y,\omega)$, and $\delta\omega=0.1$ for $N(q_x,q_y,\omega)$.
        Units are set by $t=1$.}
        \label{fig:3}
\end{figure*}

\subsection{Correlation functions of $t$-$J$ and Hubbard models}

In this section, we apply the root-$N$ CV method to the more challenging $t$-$J$ and Hubbard models on a two leg ladder geometry.
Explicit Hamiltonian expressions are given in Eqns.~\ref{eq:tJ}-\ref{eq:Hub} of Appendix~\ref{appendix_Hamis}.
We discuss $t$-$J$ model results in more detail in the Appendix~\ref{appendix_tJmodel},
and state here the main results of the analysis: the dynamical spin structure factor $S(\textbf{q},\omega)$
is practically identical in the two methods, while when considering
the dynamical charge structure factor $N(\textbf{q},\omega)$ (for a definition, see Eq.~\ref{eq:Nqw} of Appendix~\ref{appendix_Hamis}),
besides obtaining qualitative agreement, the root-$N$ provides a 
much better frequency \emph{resolution} (we recall here that in both methods
the same broadening $\eta$ was used).
We here instead focus on the Hubbard model where minor differences
in the results between the two methods can be observed when a
moderately small exponent $N$ is used in the root-$N$ CV method.
As in the $t$-$J$ case, we consider spin as well as charge dynamical structure
factors for a doped ladder ($N_{\text{el}}=0.88L$, $12\%$ hole doping)
with system size $L=50\times2$. We consider an isotropic ladder with parameters $t_x=t_y=t=1$ and $U=8t$.
Spin and charge structure factors for Hubbard ladders were already studied and discussed by us
in Refs.~\cite{re:Nocera2016,re:Nocera2017,re:Nocera2018doping,re:Kumar2019,re:tseng2022crossover}, 
where the conventional Krylov space CV method was used.
Figure \ref{fig:3} uses a maximum $m=3000$ DMRG states in both
methods for $S(\textbf{q},\omega)$ while a maximum of $m=2000$ was used for $N(\textbf{q},\omega)$.
In both cases, the minimum number of DMRG states was $m_{\text{min}}=200$, and the truncation error was kept 
smaller than $10^{-7}$.

Figure~\ref{fig:3}a-d shows the comparison for the dynamical spin structure factor $S(\textbf{q},\omega)$.
For $N=4$ the root-$N$ CV method gives results
practically identical to the CV method, and only minor quantitative differences can be observed.
For example, in the root-$N$ method, the broad two-triplon excitation band in the spin structure factor $S(q_x,q_y=0,\omega)$ appears
to be sharper than in the conventional CV method. 
In the $q_y=\pi$ compoment, instead, the main spectral features at the incommensurate wave-vector $q_x~\simeq0.88\pi$
appear slightly broader in the conventional CV method as a function of frequency, at low frequencies.
From this analysis, we conclude that even a moderately small exponent $N$ is sufficient 
to get a better converged spin spectral function using the root-$N$ CV method.

These observations are relevant
when comparing DMRG spectral data with RIXS~\cite{re:Schlappa2009} and INS~\cite{re:Notbohm2007} experiments
in the challenging ``telephone number'' cuprates, experimental data 
that recently has became available for the doped regime~\cite{re:tseng2022crossover}.

Finally, we discuss the dynamical charge structure factor, which is
of interest in RIXS measurements  of the charge-transfer band excitations in ladder cuprates.
When a Hubbard ladder is doped with holes with respect to half-filling,
we observe two branches in the $N(\textbf{q},\omega)$: the first one at low-energy
corresponds to in-band particle-hole excitations across the Fermi level. The high-energy band describes
charge-transfer electronic exitations above the Mott gap. Figure~\ref{fig:3}e-h shows that
the root-$N$ CV method provides high quality spectral data  with no appreciable shifts (downwards or upwards)
of the main features.
(Please remember that we are using the same $\eta$ for both methods.) Yet some spectral weight redistribution can be noted: spectral intensity
on the high-energy charge-transfer band appears more intense in the root-$N$ CV method compared to the conventional CV method.
We conclude that in this case, even though very good results can be obtained with a modest exponent $N$, 
one should prefer simulations with the largest possible $N$
in order to get the best results from our root-$N$ method.

\section{Discussions and Conclusions}

In this work, we have proposed a method to
compute generic spectral functions of strongly correlated Halmiltonians using
generalized correction-vectors with smaller entanglement content: the root-$N$ CV method.
The idea behind the root-$N$ CV draws inspiration in part from time dependent MPS methods,
and in part from the Chebyshev MPS approach.
The CheMPS method helps in computing spectral functions but, as was highlighted recently \cite{re:Xie2018},
while resolving accurately the low-energy part of the spectral functions, CheMPS
cannot resolve the high-energy spectrum accurately because an energy-truncation
of the Chebyshev vectors is in general required. To avoid this issue, Xie et al.~\cite{re:Xie2018}
have proposed a reorthogonalization scheme for the Chebyshev vectors (ReCheMPS). Nevertheless,
if the target frequency window for the spectral function
is chosen to be much smaller than the many body width of the system (this should be in general done to increase the frequency resolution),
an energy truncation might still be required.
There is evidence that the energy-truncation procedure severely limits
the applicability of the CheMPS or ReCheMPS methods in challenging cases as in Hubbard or $t$-$J$ models,
as in these cases it likely becomes a necessary step of the algorithm, mainly because
the many-body bandwidth is in general much larger than the spectral support of typical spectral functions.
When the energy truncation is performed, several Krylov space projections as Chebyshev recurrence steps are required,
rendering the method as computationally demanding as the conventional CV method.

Going back to the root-$N$ CV, this publication has showed that when the exponent $N$ is sufficiently large,
the root-$N$ CV performance becomes better than that of the conventional CV,
because the former method handles much less entangled correction-vectors.
In particular, we have shown evidence that in the Heisenberg and $t$-$J$ models the root-$N$ CV method
improves even the quality of the spectral functions, and provides a better frequency resolution.
Larger $N$ values in the root-$N$ CV method require more sweeping of the lattice,
but do not affect CPU times, because each sweep is faster than using smaller $N$ values.

Finally, the challenging Hubbard model requires a careful use of our root-$N$ CV method: 
while moderately small exponents $N$ give very good results for the main spectral features,
our data shows only minor differences with respect to the conventional CV method, 
which however should be taken into account when high-precision experimental results are available.

We believe that root-$N$ correction-vector DMRG will
become a much used, not only
when high precision spectral data is sought, but also when
high performance is required, performance better than the computationally expensive
conventional CV method.

The root-$N$ method should also facilitate
high precision spectral function calculations in finite width cylinders, where better computational methods are currently needed. 
These cylinders try to approach the two-dimensional models
that are at the frontier of what DMRG can do, and they need
a very large computational effort to simulate. 

\section*{Acknowledgments}
A. Nocera acknowledges support from the Max Planck-UBC-UTokyo Center for Quantum Materials
and Canada First Research  Excellence Fund (CFREF) Quantum Materials and Future Technologies Program
of the Stewart Blusson Quantum Matter Institute (SBQMI), and the Natural Sciences and Engineering
Research Council of Canada (NSERC). This work used computational resources and services provided
by Compute Canada and Advanced Research Computing at the University of British Columbia.
G.A. was partially supported by the Scientific Discovery through
Advanced Computing (SciDAC) program funded by U.S.
DOE, Office of Science, Advanced Scientific Computing
Research and BES, Division of Materials Sciences and
Engineering.

\bibliography{biblio}
\newpage

\appendix
\section{MPS algorithm to build the root-$N$ correction-vector}\label{appendix_MPSalgo}
\begin{figure}
        \centering
        \includegraphics[width=0.48\textwidth]{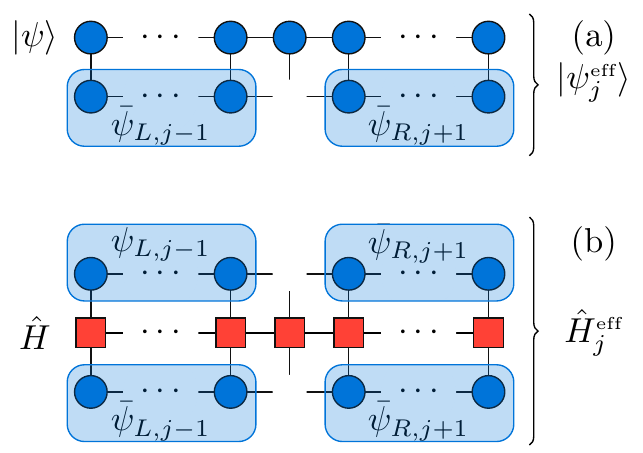}
        \caption{\textbf{Effective local state vector and Hamiltonian.}
	Panel (a): Effective state $|\psi^{\text{eff}}_j\rangle$ obtained by projecting the MPS by the maps
	$\bar{\psi}_{L,j-1}\equiv (M^{\dag}_1,...,M^{\dag}_{j-1})$ and
	$\bar{\psi}_{R,j+1} \equiv (M^{\dag}_{j+1},...,M^{\dag}_L)$. If $|\psi\rangle$ is a mixed-canonical
	MPS representation, then simply $|\psi^{\text{eff}}_j\rangle=M_{j}$.
	Panel (b): Effective (one-site) Hamiltonian obtained by projecting $\hat{H}$ using the maps
	$\{\bar{\psi}_{L,j-1},\psi_{L,j-1},\bar{\psi}_{R,j+1},\psi_{R,j+1}\}$ defined above.
	Analogous definitions can be given in the two-site case.}
        \label{fig:App}
\end{figure}

		Let us introduce a Matrix Product State representing
		the ground state of the system for $L$ sites and open boundary conditions
		(we use a notation similar to Ref.~\cite{re:Paeckel2019})
		\begin{equation}
			|\psi\rangle = \sum\limits_{\substack{\sigma_1...\sigma_L\\m_0...m_L}} M^{\sigma_1}_{1;m_0,m_1}...M^{\sigma_L}_{L;m_{L-1},m_L} |\sigma_1...\sigma_L\rangle,
		\end{equation}
		where $m_{j}$ are the bond dimensions or virtual indices
		(with $m_0$ and $m_L$ 1-dimensional dummy indices), and $\sigma_{j}$ represent
		the physical indices of the many-body state of the system. Formally, let us define
		the tensors $\bar{\psi}_{L,j-1}\equiv (M^{\dag}_1,...,M^{\dag}_{j-1})$ and
		$\bar{\psi}_{R,j+1} \equiv (M^{\dag}_{j+1},...,M^{\dag}_L)$ which constitute a left and right map, respectively,
		from the joint Hilbert space on sites 1 through $j-1$ onto the bond space $m_{j-1}$,
		and from the joint Hilbert space on sites $j+1$ through $L$ onto the bond space $m_j$.
		If we apply these maps to the MPS $|\psi\rangle$, we can obtain the effective
		state at site $j$, $|\psi^{\text{eff}}_j\rangle$; see Fig.~\ref{fig:App}a.
		When $|\psi\rangle$ is in a MPS mixed-canonical form,
		$|\psi^{\text{eff}}_j\rangle$ equals the 3-rank tensor $M_{j,m_{j-1}m_j}$ in the MPS at site $j$,
		which is often interpreted as a vector of dimensions $(d_{j} \times m_{j-1} \times m_{j})$, where 
		$d_{j}$ is the local physical Hilbert space dimension.
		Similarly, the Hamiltonian $\hat{H}$, in matrix product operator (MPO) form,
		acts between the maps defined above
		(and their conjugates, $\psi_{L,j-1}$ $\psi_{R,j+1}$; see Fig.~\ref{fig:App}b)
		to yield an effective single site Hamiltonian $\hat{H}^{\text{eff}}_{j}$.
		This procedure can also be defined in the space of two-sites.
		A computer program never needs to explicitly construct $\hat{H}^{\text{eff}}_{j}$,
		but only evaluates its action on $|\psi^{\text{eff}}_j\rangle$.

		Using $|\psi^{\text{eff}}_j\rangle$ and  $\hat{H}^{\text{eff}}_{j}$, we construct three local MPS tensors.
		The first one is obtained by applying the operator $\hat{O}_j$ on $|\psi^{\text{eff}}_j\rangle$,
		yielding $|\phi\rangle = \hat{O}_j|\psi\rangle$. The MPS $|\phi\rangle$ has all the
		tensors equal to those of $|\psi\rangle$ except for the one at site $j$, $M^{\prime}_{j;m_{j-1},m_{j}}$
		\begin{equation}
			M^{\prime}_{j;m_{j-1},m_{j}} = \sum_{\sigma^{\prime}_{j}} O^{\sigma_j \sigma^{\prime}_j}_{j}M^{\sigma^{\prime}_j}_{j;m_{j-1},m_{j}}
		\end{equation}
		We then construct the (real and imaginary part of the) root-$N$ correction-vector by Krylov space decomposition
		of the Hamiltonian $\hat{H}^{\text{eff}}_{j}$
		\begin{widetext}
                \begin{equation}
			[X(\omega+i\eta)]^{\sigma_j}_{j;m_{j-1},m_{j}} = \sum\limits_{\substack{l,l^{\prime}nn^{\prime}\\ \sigma^{\prime}_j,m^{\prime}_{j-1},m^{\prime}_{j} } } T^{\dag}_{l,\sigma_j,m_{j-1},m_{j}} P^{\dag}_{ln} \frac{1}{[\omega-\epsilon^{\text{eff}}_j \delta_{n n^{\prime}}+E_g+i\eta]^{1/N}} P_{n^{\prime}l^{\prime}}T_{l,\sigma^{\prime}_j,m^{\prime}_{j-1},m^{\prime}_{j}} M^{\sigma^{\prime}_j}_{j;m^{\prime}_{j-1},m^{\prime}_{j}}
                \end{equation}
		\end{widetext}
		where $T_{l,\sigma_j,m_{j-1},m_{j}}$ tridiagonalizes $\hat{H}^{\text{eff}}_{j}$,
		$\hat{H}^{\text{Tridiag,eff}}_{j}=T^{\dag} \hat{H}^{\text{eff}}_j T$, to the smaller Krylov space spanned
		by the index $l$, $\text{dim}[l]<<d_j \times\text{dim}[m_{j-1}] \times \text{dim}[m_{j}]$.
		$P_{ln}$ diagonalizes
		$\hat{H}^{\text{Tridiag,eff}}_{j}$, $\hat{H}^{\text{diag,eff}}_{j} = P^{\dag}\hat{H}^{\text{Tridiag,eff}}_{j}P$,
		where $\epsilon^{\text{eff}}_j$ are the eigenvalues of $\hat{H}^{\text{diag,eff}}_{j}$.
                How is the Krylov space tridiagonalization of $\hat{H}^{\text{eff}}_{j}$ stopped?
		In practice, we compare the lowest eigenvalue of $\hat{H}^{\text{diag,eff}}_{j}$,
		$\epsilon_{\text{min}}=\{\epsilon^{\text{eff}}_j [k]\}_{\text{min}}$
		at iteration $k$ and $k+1$, and exit the loop when the error breaks below a certain threshold.
		We set this error to $\epsilon_{\text{Tridiag}}=10^{-9}$ in order to avoid the proliferation of
		Krylov vectors (and thus Lanczos iterations), and their reorthogonalizations.
		In general, the three states $|\phi\rangle,~|X^{\text{Re}}\rangle,~|X^{\text{Im}}\rangle$
		will be represented in a bad basis of the environments
		$\psi_L$ and $\psi_R$ which are optimized to represent the original state $|\psi\rangle$.
		To expand these bases, we use state-averaging of the four states, which is equivalent to
		targeting more than one state in conventional DMRG language.
		In MPS language, as explained in Ref.~\cite{re:YangM2020}, the state-
		averaging is done by creating an extra index which labels the
		states involved. One formally considers an expanded MPS representing a mixed state
		\begin{widetext}
		\begin{align*}
			&\begin{pmatrix}|\psi\rangle\\|\phi\rangle\\|X^{\text{Re}}\rangle\\|X^{\text{Im}}\rangle\end{pmatrix} =
				\sum_{\sigma_1...\sigma_L} A^{\prime,\bar{\sigma}_1}_1 ...
				C^{\prime,\bar{\sigma}_j}_{j} ... B^{\prime,\bar{\sigma}_{L}}_L|\sigma_1...\sigma_L\rangle\nonumber\\
				&=\sum_{\sigma_1...\sigma_L}
				\begin{pmatrix} A^{\sigma_1}_1[\psi]& 0& 0& 0\\ 0& A^{\sigma_1}_1[\phi]& 0& 0\\
					0& 0& A^{\sigma_1}_1[X^{\text{Re}}]& 0\\ 0& 0& 0& A^{\sigma_1}_1[X^{\text{Im}}]\end{pmatrix}
					... \begin{pmatrix}C^{\sigma_j}_j[\psi]\\C^{\sigma_j}_j[\phi]\\C^{\sigma_j}_j[X^{\text{Re}}]\\C^{\sigma_j}_j[X^{\text{Im}}]\end{pmatrix} ... \begin{pmatrix} B^{\sigma_L}_L[\psi]& 0& 0& 0\\ 0& B^{\sigma_L}_L[\phi]& 0& 0\\
					0& 0& B^{\sigma_L}_L[X^{\text{Re}}]& 0\\ 0& 0& 0& B^{\sigma_L}_L[X^{\text{Im}}]\end{pmatrix}
					|\sigma_1...\sigma_L\rangle,
		\end{align*}
		\end{widetext}
		where $C^{\prime,\bar{\sigma}_j}_{j;m^{\prime}_{j-1},m^{\prime}_j}$ has four
		components (representing the four \emph{targeted} vectors) and it has extended bond dimensions
		$m^{\prime}_{j-1}=m^{[\psi]}_{j-1}+m^{[\phi]}_{j-1}+m^{[X^{\text{Re}}]}_{j-1}+m^{[X^{\text{Im}}]}_{j-1}$,
		$m^{\prime}_{j}=m^{[\psi]}_{j}+m^{[\phi]}_{j}+m^{[X^{\text{Re}}]}_{j}+m^{[X^{\text{Im}}]}_{j}$.
		Here, the notation in terms of $A$ and $B$ tensors underlines a mixed-canonical representation of all the MPSs.
		By SVD compression, one has
		\begin{equation}\label{eq:App}
			C^{\prime,\sigma_j}_j = U_j^{\prime,\sigma_j}S_j^{\prime}V_j^{\prime,\dag}.
		\end{equation}
		As in conventional DMRG, one can also introduce different weights in the direct sum and
		perform a SVD of the weighted sum of the reduced density matrix $\rho^{\prime}=\sum^{3}_{k=0}w_k \rho_{k}$
		Once this procedure is performed at site $j$, one can proceed updating all the tensors at site $j+1$. In formulas,
		\begin{equation}
			C^{\prime}_{j+1} = \begin{pmatrix}C^{\sigma_{j+1}}_{j+1}[\psi]\\C^{\sigma_{j+1}}_{j+1}[\phi]\\C^{\sigma_{j+1}}_{j+1}[X^{\text{Re}}]\\C^{\sigma_{j+1}}_{j+1}[X^{\text{Im}}]\end{pmatrix} =  \begin{pmatrix}U^{[\prime,\dag,\sigma_{j}]}_jC^{\sigma_{j}}_j[\psi]B^{\sigma_{j+1}}_{j+1}[\psi]\\U^{[\prime,\dag,\sigma_{j}]}_{j}C^{\sigma_{j}}_j[\phi]B^{\sigma_{j+1}}_{j+1}[\phi]\\U^{[\prime,\dag,\sigma_{j}]}_jC^{\sigma_{j}}_j[X^{\text{Re}}]B^{\sigma_{j+1}}_{j+1}[X^{\text{Re}}]\\U^{[\prime,\dag,\sigma_{j}]}_jC^{\sigma_{j}}_j[X^{\text{Im}}]B^{\sigma_{j+1}}_{j+1}[X^{\text{Im}}]\end{pmatrix},
		\end{equation}
		where the $U^{[\prime,\dag,\sigma_{j}]}_j$ from Eq.~\ref{eq:App} 
		is common to all the four vectors. After sweeping back and forth through the lattice,
		a \emph{good} representation of the correction-vectors is obtained.

\section{Hamiltonians and dynamical structure factors in real space}\label{appendix_Hamis}

The Heisenberg Hamiltonian on a two leg ladder geometry with open boundary conditions and size 
$L=L_x\times 2$ is defined as
\begin{equation}\label{eq:Heis}
	H_{\rm Heis} = J_x \sum^{L_x-1}_{i=1;\gamma=1,2} \textbf{S}_{i,\gamma} \cdot \textbf{S}_{i+1,\gamma} 
	+ J_y \sum^{L_x}_{i=1} \textbf{S}_{i,1} \cdot \textbf{S}_{i,2},
\end{equation}
where $\textbf{S}_{i,\gamma}\equiv\{S^{x}_{i,\gamma},S^{y}_{i,\gamma},S^{z}_{i,\gamma}\}$ describe the spin 1/2 
operators on site $i$ and ladder leg $\gamma$. Similarly, the $t$-$J$ Hamiltonian is defined as
\begin{align}\label{eq:tJ}
	H_{t-J} &= -t_x\sum^{L_x-1}_{i=1;\gamma=1,2;\sigma} \Big(c^{\dag}_{i,\gamma,\sigma} c_{i+1,\gamma,\sigma}+{\rm h.c.}\Big)\nonumber\\
	&-t_y\sum^{L_x}_{i=1;\sigma} \Big(c^\dag_{i,1,\sigma} c_{i,2,\sigma} +{\rm h.c.}\Big)\nonumber\\
	&+J_x \sum^{L_x-1}_{i=1;\gamma=1,2} \Bigg(\textbf{S}_{i,\gamma} \cdot \textbf{S}_{i+1,\gamma}-\frac{n_{i,\gamma}n_{i+1,\gamma}}{4}\Bigg)\nonumber\\
	&+ J_y \sum^{L_x}_{i=1} \Bigg(\textbf{S}_{i,1} \cdot \textbf{S}_{i,2} - \frac{n_{i,1}n_{i,2}}{4}\Bigg),
\end{align}
where $c^\dag_{i,\gamma,\sigma}$ ($c_{i,\gamma,\sigma}$)
is the electron creation (annihilation) operator on site $i$, ladder leg $\gamma$ 
with spin polarization $\sigma$, while $n_{i,\gamma} = \sum_{\sigma}c^\dag_{i,\gamma,\sigma}c_{i,\gamma,\sigma}$ is the electron
number operator. Finally, the Hubbard Hamiltonian is 
\begin{align}\label{eq:Hub}
	&H_{\rm Hub} = -t_x\sum^{L_x-1}_{i=1;\gamma=1,2;\sigma} \Big(c^{\dag}_{i,\gamma,\sigma} c_{i+1,\gamma,\sigma}+{\rm h.c.}\Big)\nonumber\\
        &-t_y\sum^{L_x}_{i=1;\sigma} \Big(c^\dag_{i,1,\sigma} c_{i,2,\sigma} +{\rm h.c.}\Big)+U\sum^{L_x}_{i=1;\gamma=1,2} n_{i,\gamma,\uparrow}n_{i,\gamma,\downarrow}.
\end{align}
The spin structure factor $S(\textbf{q},\omega)$ with $\textbf{q}\equiv(q_x,q_y)$ can be defined as 
\begin{align}\label{eq:Sqw}
	S(\textbf{q},\omega) = \frac{1}{2L_x}\sum^{L_x}_{j=1;\gamma=1,2} e^{i [q_x(j-i) + q_y(\gamma-\gamma^{\prime})]}\times\nonumber\\
	\langle\psi|\textbf{S}_{j,\gamma}\frac{1}{\omega-H+i\eta}\textbf{S}_{i,\gamma^{\prime}}|\psi\rangle,
\end{align}
where the \emph{center} point is chosen in the middle of the leg 1, $(i,\gamma^{\prime})=(L_x/2,1)$.
Analogously, the charge structure factor is 
\begin{align}\label{eq:Nqw}
        N(\textbf{q},\omega) = \frac{1}{2L_x}\sum^{L_x}_{j=1;\gamma=1,2} e^{i [q_x(j-i) + q_y(\gamma-\gamma^{\prime})]} \times\nonumber\\
	\langle\psi|\delta n_{j,\gamma}\frac{1}{\omega-H+i\eta}\delta n_{i,\gamma^{\prime}}|\psi\rangle,
\end{align}
where $\delta n_{j,\gamma}\equiv n_{j,\gamma}-\langle\psi|n_{j,\gamma}|\psi\rangle$, 
where $|\psi\rangle$ is the ground state of the system.

\section{Dynamical Structure factors for the $t$-$J$ model on a two leg ladder geometry}\label{appendix_tJmodel}

This section of the appendix applies the root-$N$ CV method to a $t$-$J$ model on a two-leg ladder geometry, and compares
the results obtained to the conventional CV approach. We calculate both spin and charge dynamical structure
factors for a doped ladder with $N_{\text{el}}=0.88L$, corresponding to $12\%$ hole doping, and with
lattice size $L=50\times2$. In this case, we use a maximum $m=1200$ DMRG states for both methods
(and a minimum $m_{\text{min}}=200$), in order to keep the truncation error below $10^{-8}$.

\begin{figure*}
\centering
\begin{minipage}{.45\textwidth}
  \centering
  \includegraphics[width=\columnwidth]{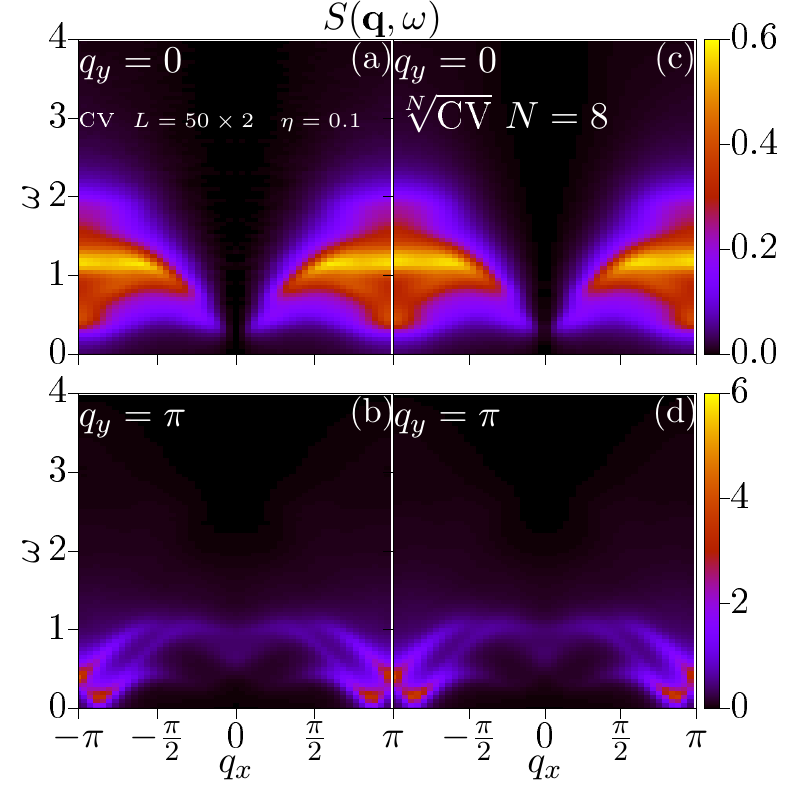}
\end{minipage}%
\begin{minipage}{.45\textwidth}
  \centering
  \includegraphics[width=\columnwidth]{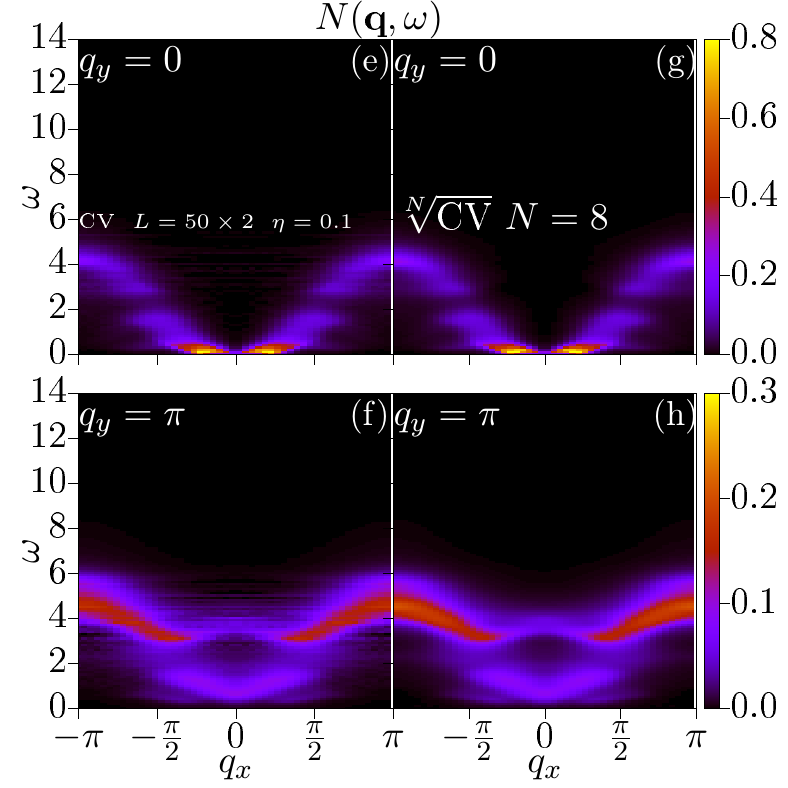}
\end{minipage}
        \caption{\textbf{Dynamical structure factors for a $t$-$J$ two-leg ladder.}
        Panels (a)-(b) ((e)-(f)) report the $q_y=0,\pi$ components of the $S(q_x,q_y,\omega)$ ($N(q_x,q_y,\omega)$)
        using the standard Krylov space correction-vector approach.
        A ladder with $t_y=t_x=t=1$, $J_x=J_y=0.5t$ is simulated.
        Length is $L=50\times2$, number of electrons $N_{\text{el}}=0.88L$, broadening $\eta=0.1$,
        and resolution step $\delta\omega=0.1$ (units are set by $t=1$).
        Panels (c)-(d) ((g)-(h)) report the $q_y=0,\pi$ components of the $S(q_x,q_y,\omega)$ ($N(q_x,q_y,\omega)$) using the root-$N$
        Krylov space correction-vector using $N=8$.}
	\label{fig:5}
\end{figure*}

Figure~\ref{fig:5}a-d shows the comparison for the dynamical spin structure factor $S(\textbf{q},\omega)$.
We note that for $N=8$ the root-$N$ CV method yields results
that are practically identical to those obtained with the CV method.
Yet the root-$N$ method  provides a much better frequency resolution for the
more challenging dynamical charge structure factor $N(\textbf{q},\omega)$, where
we also obtain quantitative agreement.

\newpage
	
\cleardoublepage
\newpage
\onecolumngrid

\section*{Supplementary Material: Computational details to reproduce the DMRG results}

Here we provide instructions on how to reproduce the DMRG results used in the main text. 
The results reported in this work were obtained with DMRG++ versions 6.01 and PsimagLite versions 3.01. 
The DMRG++ computer program~\cite{re:Alvarez0209} can be obtained with:
\begin{verbatim}
git clone https://github.com/g1257/dmrgpp.git
git clone https://github.com/g1257/PsimagLite.git
\end{verbatim}
The main dependencies of the code are BOOST and HDF5 libraries.
To compile the program:
\begin{verbatim}
cd PsimagLite/lib; perl configure.pl; make
cd ../../dmrgpp/src; perl configure.pl; make
\end{verbatim}

The DMRG++ documentation  can  be  found  at  \verb! https://g1257.github.io/dmrgPlusPlus/manual.html! or  can  be  obtained  by doing 
\verb!cd dmrgpp/doc; make manual.pdf!. In the description of the DMRG++ inputs below,
we follow very closely the description in the supplemental material of Ref.~\cite{re:Scheie2021witnessing}, where similar calculations were performed.

The spectral function results for the Heisenberg model on the two leg ladder geometry can be reproduced as follows. We first run 
\verb!./dmrg -f inputGS.ain -p 12! to obtain the ground state wave-function and ground state energy with 12 digit precision using the \verb!-p 12! option. The \verb!inputGS.ain! has the form
\begin{verbatim}
##Ainur1.0
TotalNumberOfSites=100;
NumberOfTerms=2;

### 1/2(S^+S^- + S^-S^+) 
gt0:DegreesOfFreedom=1;
gt0:GeometryKind="ladder";
gt0:GeometryOptions="ConstantValues";
gt0:dir0:Connectors=[1.0];
gt0:dir1:Connectors=[2.0];
gt0:LadderLeg=2;

### S^zS^z part
gt1:DegreesOfFreedom=1;
gt1:GeometryKind="chain";
gt1:GeometryOptions="ConstantValues";
gt1:dir0:Connectors=[1.0];
gt1:dir1:Connectors=[2.0];
gt1:LadderLeg=2;

Model="Heisenberg";
HeisenbergTwiceS=2;
SolverOptions="twositedmrg,useComplex";
InfiniteLoopKeptStates=200;
FiniteLoops=[[49, 2000, 0],
[-98, 2000, 0],
[98, 2000, 0],
[-98, 2000, 0],
[98, 2000, 0]];

# Keep a maximum of 1000 states, but allow SVD truncation with 
# tolerance 1e-12 and minimum states equal to 200
TruncationTolerance="1e-12,200";
# Symmetry sector for ground state S^z_tot=0
TargetSzPlusConst=50    
OutputFile="dataGS_L100";
\end{verbatim}

The parameter TargetSzPlusConst should be equal $Sz+L/2$, where $Sz$ is the targeted $Sz$ sector and $L$ is the system size. 
The next step is to calculate dynamics for the $S(\textbf{q},\omega)$ spectral function using the saved ground state as an input. 
It is convenient to do the dynamics run in a subdirectory \verb!Sqw!, so \verb!cp inputGS.ain Sqw/inputSqw.ado! and add/modify the following lines in \verb!inputSqw.ado!
\begin{verbatim}
# The finite loops now start from the final loop of the gs calculation. 
# Total number of finite loops equal to N, here N=8
FiniteLoops=[
[-98, 2000, 2],[98, 2000, 2],
[-98, 2000, 2],[98, 2000, 2],
[-98, 2000, 2],[98, 2000, 2],
[-98, 2000, 2],[98, 2000, 2]];

# The exponent in the root-N CV method
CVnForFraction=8;

# Solver options should appear on one line, here we have two lines because of formatting purposes
SolverOptions="calcAndPrintEntropies,useComplex,twositedmrg,
                TargetingCVEvolution,restart,fixLegacyBugs,minimizeDisk";
CorrectionA=0;

# RestartFilename is the name of the GS .hd5 file (extension is not needed) 
RestartFilename="../dataGS_L100";

# The weight of the g.s. in the density matrix
GsWeight=0.1;
# Legacy, set to 0
CorrectionA=0;
# Fermion spectra has sign changes in denominator. 
# For boson operators (as in here) set it to 0
DynamicDmrgType=0;
# The site(s) where to apply the operator below. Here it is the center site.
TSPSites=[48];
# The delay in loop units before applying the operator. Set to 0
TSPLoops=[0];
# If more than one operator is to be applied, how they should be combined.
# Irrelevant if only one operator is applied, as is the case here.
TSPProductOrSum="sum";
# How the operator to be applied will be specified
string TSPOp0:TSPOperator=expression;
# The operator expression
string TSPOp0:OperatorExpression="sz";
# How is the freq. given in the denominator (Matsubara is the other option)
CorrectionVectorFreqType="Real";
# This is a dollarized input, so the 
# omega will change from input to input.
CorrectionVectorOmega=$omega;
# The broadening for the spectrum in omega + i*eta
CorrectionVectorEta=0.1;
# The algorithm
CorrectionVectorAlgorithm="Krylov";
#The labels below are ONLY read by manyOmegas.pl script
# How many inputs files to create
#OmegaTotal=60
# Which one is the first omega value
#OmegaBegin=0.0
# Which is the "step" in omega
#OmegaStep=0.1
# Because the script will also be creating the batches, 
# indicate what to measure in the batches
#Observable=sz    
\end{verbatim}

Notice that the main change with respect of a standard CV method input is 
given by the option \verb!TargetingCVEvolution! in the SolverOptions
instead of \verb!CorrectionVectorTargeting!, and the addition of the line \verb!CVnForFraction=8;!
We note also that the number of finite loops must be at least equal to the number 
equal to the exponent in the root-$N$ CV method.
As in the standard CV approach, all individual  inputs  
(one  per $\omega$ in  the  correction  vector  approach) 
can  be  generated  and  submitted  using  the \verb!manyOmegas.pl! script which can be found in the \verb!dmrgpp/src/script! folder:
\begin{verbatim}
perl manyOmegas.pl inputSqw.ado BatchTemplate.pbs <test/submit>.
\end{verbatim}
It is recommended to run with \verb!test! first to verify correctness, before running with \verb!submit!.
Depending on the machine and scheduler, the \verb!BatchTemplate! can be e.g. a PBS or SLURM script. 
The key is that it contains a line \verb!./dmrg -f $$input "<X0|$$obs|P2>" -p 12! which allows \verb!manyOmegas.pl! to fill in the appropriate input for each generated job batch. 
After all outputs have been generated,
\begin{verbatim}
perl procOmegas.pl -f inputSqw.ado -p
perl pgfplot.pl
\end{verbatim}
can be used to process and plot the results (these scripts are also given in 
\verb!dmrgpp/src/script! folder).

For the $t$-$J$ model calculations of the main text, the following substitutions should be applied in the
ground state input
\begin{verbatim}
NumberOfTerms=4;

###  Kinetic term c^+ c + h.c.
gt0:DegreesOfFreedom=1;
gt0:GeometryKind="ladder";
gt0:GeometryOptions="ConstantValues";
gt0:dir0:Connectors=[-1.0];
gt0:dir1:Connectors=[-1.0];
gt0:LadderLeg=2;

### 1/2(S^+S^- + S^-S^+)
gt1:DegreesOfFreedom=1;
gt1:GeometryKind="ladder";
gt1:GeometryOptions="ConstantValues";
gt1:dir0:Connectors=[0.5];
gt1:dir1:Connectors=[0.5];
gt1:LadderLeg=2;

### S^zS^z part
gt2:DegreesOfFreedom=1;
gt2:GeometryKind="chain";
gt2:GeometryOptions="ConstantValues";
gt2:dir0:Connectors=[0.5];
gt2:dir1:Connectors=[0.5];
gt2:LadderLeg=2;

### density-density n*n
gt3:DegreesOfFreedom=1;
gt3:GeometryKind="ladder";
gt3:GeometryOptions="ConstantValues";
gt3:dir0:Connectors=[-0.125];
gt3:dir1:Connectors=[-0.125];
gt3:LadderLeg=2;

Model="TjMultiOrb";
Orbitals=1;
potentialV=[0.0,...];

# Total number of electrons is 88, such that N_el = 0.88*L
TargetElectronsUp=44;
TargetElectronsDown=44;

\end{verbatim}

The input should then be run as \verb!./dmrg -f inputGS.ain -p 12 <gs|n|gs>! where the local density at the center of the 
ladder should be saved as it will be needed later in the dynamics input.
For the dynamical charge structure factor $N(\textbf{q},\omega)$ the only additional modification in the \verb!inputNqw.ado! file is
\begin{verbatim}
string TSPOp0:OperatorExpression="n+(-0.8955)*identity";
#Observable=n
\end{verbatim}
where the ground-state density at the center site $\langle\psi|\hat{n}_{48}|\psi\rangle$ 
needs to be subtracted to avoid a delta function peak at zero frequency and momentum.

For the Hubbard model, the modifications are
\begin{verbatim}
NumberOfTerms=1;

###  Kinetic term c^+ c + h.c.
DegreesOfFreedom=1;
GeometryKind="ladder";
GeometryOptions="ConstantValues";
dir0:Connectors=[-1.0];
dir1:Connectors=[-1.0];
LadderLeg=2;

Model="HubbardOneBand";

hubbardU=[8.0,...];
potentialV=[0.0,...];

# Total number of electrons is 88, such that N_el = 0.88*L
TargetElectronsUp=44;
TargetElectronsDown=44;

\end{verbatim}
Finally, for the dynamical charge structure factor calculations, the only additional modification in the \verb!inputNqw.ado! file is
\begin{verbatim}
string TSPOp0:OperatorExpression="n+(-0.8995)*identity";
#Observable=n
\end{verbatim}
	
\end{document}